\documentclass[aps,prl,twocolumn,superscriptaddress,amsfont,amssymb,amsmath,nofootinbib,showpacs,balancelastpage]{revtex4-1}
\usepackage{float}
\usepackage{epsfig}
\usepackage{graphicx}
\usepackage{dcolumn}
\usepackage{bm}
\usepackage{color}
\usepackage{tabularx}
\usepackage{cleveref}

\newcommand{\lsim}{\mbox{\raisebox{-.9ex}{~$\stackrel{\mbox{$<$}}{\sim}$~}}}

\newcommand{\Od}{{\mathcal O}}

\def\thebiblio#1{
\begin{center}\bf \large References
\end{center}
\list
{[\arabic{enumi}]}{\settowidth\labelwidth{#1.}\leftmargin\labelwidth
 \advance\leftmargin\labelsep
 \usecounter{enumi}}
 \def\newblock{\hskip .11em plus .33em minus -.07em}
 \sloppy
 \sfcode`\.=1000\relax}

\begin{document}

\preprint{}
\title{Higgs effective potential in a perturbed Robertson-Walker background
}

\author{Antonio L. Maroto}
\email{maroto@ucm.es}
\affiliation{Departamento de F\'{\i}sica Te\'orica, Universidad Complutense de Madrid, 28040 
Madrid, Spain}
\author{Francisco Prada}
\email{f.prada@csic.es}
\affiliation{Instituto de F{\'i}sica Te{\'o}rica, (UAM/CSIC), Universidad Aut{\'o}noma de Madrid, Cantoblanco, E-28049 Madrid, Spain}
\affiliation{Campus of International Excellence UAM+CSIC, Cantoblanco, E-28049 Madrid, Spain}
\affiliation{Instituto de Astrof{\'i}sica de Andaluc{\'i}a (CSIC), Glorieta de la Astronom{\'i}a, E-18080 Granada, Spain}

\date{\today}

\begin{abstract}
We calculate the one-loop effective potential of a scalar field in a Robertson-Walker
background with scalar metric perturbations. A 
complete set of orthonormal solutions of the perturbed equations is obtained by using the 
adiabatic approximation for  comoving observers. After analyzing the problem of renormalization in inhomogeneous backgrounds, we get the explicit contribution of metric 
perturbations to the effective potential.  We apply these results to the Standard Model Higgs field and evaluate
the effects of metric perturbations on the Higgs mass and on its vacuum expectation value. 
Space-time variations are found, which are proportional to the gravitational
slip parameter, with a typical amplitude of the order of $\Delta\phi/\phi\simeq 10^{-11}$  on cosmological scales. 
We also discuss possible  astrophysical signatures in the Solar System and in the Milky Way that could open new possibilities to explore the symmetry breaking sector of the electroweak interactions.

\end{abstract}

\maketitle

\section{Introduction}
The recent discovery \cite{Higgs1,Higgs2} of a scalar resonance with mass $m_H\simeq 125$ GeV compatible with the 
Standard Model (SM) Higgs particle is giving support to the idea that the 
symmetry breaking sector  (SBS) of the electroweak interactions can be 
described by a simple model based on a 
single electroweak scalar doublet. In the near future, the Large Hadron Collider (LHC) detectors ATLAS and CMS 
will be able to improve the precision on the measurements not only of  the Higgs mass 
but also of the Higgs branching ratios \cite{Higgsprop}, thus allowing to explore also the
interactions of the Higgs particle with the fermionic and gauge sectors. 
However, certain
fundamental aspects of the SBS, such as the shape of the symmetry breaking potential, 
which would allow to discriminate
the minimal model from other alternatives, are much more difficult to constrain. 
As a matter of fact, because of the small Higgs pair production cross-section, LHC will not be able to measure the Higgs self couplings, unless the 
luminosity is increased far beyond the design value.
In this respect, it would be a matter of the utmost importance to investigate alternative ways to probe the SBS and search for new particles, 
together with the analysis of Higgs rare decays.\cite{future}. 

Apart from colliders, the other natural scenario
in which the electroweak SBS could be explored is in the field of cosmology. However
current observations are not able to probe the extremely high redshifts
corresponding to the electroweak phase transition. At late times, the large mass of
the Higgs, compared to typical cosmological energy scales, makes it extremely 
difficult to find signals of the Higgs field from astrophysical observations.
However, we know that although for a scalar field the tree-level potential is not modified by the space-time metric, 
quantum fluctuations of any field are sensitive to the background geometry \cite{Mann}. Thus, 
on general grounds, we expect that the universe expansion \cite{Maggiore,Sola}, and 
what is more interesting from the observational point of view,  metric perturbations will imprint distinctive signals in
the one-loop contribution to the effective potential. Since this potential determines the vacuum 
expectation value (VEV) of the Higgs field, the fluctuations pattern could also be  present
in the particle masses themselves, thus opening the possibility of testing the 
Higgs sector from the variation of fundamental constants.

Two different approaches have been used in the literarature in order to calculate  the one-loop effective potential in curved space-times. On one hand we have 
the so called Schwinger-de Witt expansion \cite{SdW,normal,Shapiro}.
This is a covariant and local expansion of the effective action
in derivatives of the background fields over the mass of the quantum
fluctuation. This expression can be obtained \cite{normal} by expanding the background 
metric around a given space-time point using coordinates associated
to a free-falling observer (Riemann normal coordinates), and
it is therefore only valid in a normal neighborhood of the expansion point.
This method can be improved by making use of the renormalization group equations \cite{RGE}. 
On the other hand, we have the adiabatic mode expansion which is valid for slowly 
varying background metrics. It is not manifestly covariant but the results
are valid globally.  This approach has been followed in \cite{ParkerFulling,Ringwald,Hu,Maggiore} for homogeneous and isotropic Robertson-Walker
backgrounds, for anisotropic Bianchi I cosmologies in \cite{HuangBianchi}, and in inhomogeneous space-times as those in which we are interested  
\cite{Huanginho,Albareti}.  In \cite{Schwarzschild}, the one-loop effective potential of the Higgs field
was calculated on a Schwarzschild background. The results showed a dependence of the potential on the space-time point which implies a
shift of all the particle masses near the black hole.  The possibility that quantum effects on non-trivial gravitational background could lead
to the violation of local position invariance has been discussed also in \cite{Mann}.

In this work we will focus on the modifcation of the effective potential 
on cosmological scales, i.e. we would like to  compare the effective potential in space-time
points separated by arbitrary distances, which can be larger than 
the curvature radius of spatial sections. Therefore, the 
second approach is more appropriate for our purposes. Unlike \cite{Huanginho}, 
 we have been able to obtain explicit mode solutions in the Wentzel-Kramers-Brillouin (WKB) 
approximation
without adopting the early time approximation, and without additional Taylor
expansions of the background metric.

This paper is organized as follows. First, we will  review the standard calculation of the effective potential in Minkowski space-time using canonical quantization and 
summing over the Fourier modes. Then we follow the same approach in 
perturbed Robertson-Walker backgrounds. We obtain a complete set of orthonormal
mode solutions and calculate the corresponding homogeneous and inhomogeneous 
contributions to the Higgs effective potential. We discuss the problem of renormalization and estimate the variations in the Higgs VEV and mass generated by metric perturbations. The results show a dependence on
the difference between the two metric scalar potentials (gravitational slip). Finally, possible observational signatures of the variation induced in the particle masses are discussed.

\section{One-loop effective potential in Minkowski space-time}
In order to present the calculation techniques that will be used along the paper, 
firstly  we will briefly review the standard derivation of the scalar one-loop effective potential
in flat space-time using canonical quantization methods. 

Let us consider the action for a scalar field with potential
$V(\phi)$:
\begin{eqnarray}
S[\phi]=\int d^4x \left(\frac{1}{2}\eta^{\mu\nu}\partial_\mu\phi\partial_\nu\phi-V(\phi)\right).
\end{eqnarray}
The corrresponding equations of motion read
\begin{eqnarray}
\Box \, \phi +V'(\phi) = 0 \, ,
\label{KG}
\end{eqnarray}
where prime denotes the derivative with respect to the argument. The field $\phi$ can be decomposed
into a homogeneous  $\hat \phi(t)$ and an inhomogeneous component $\delta \phi(t,{\bf x})$ i.e.,
\begin{eqnarray}
\phi(t,{\bf x})=\hat \phi(t)+\delta \phi(t,{\bf x}) ,
\end{eqnarray}
where the spatial average of  $\delta \phi(t,{\bf x})$ vanishes.
By substituting back in (\ref{KG}), expanding the potential around the 
homogeneous value and averaging over space, we get to the lowest order: 
\begin{eqnarray}
\Box \, \hat\phi +V'(\hat\phi)+\frac{1}{2}V'''(\hat\phi)\langle \delta \phi^2\rangle = 0 \, ,
\label{flucteq}
\end{eqnarray}
where cubic $\langle \delta \phi^3\rangle$ and higher-order terms have been neglected. 
When quantizing the 
fluctuation field $\delta \phi$, the combination of the last two terms
will give rise to the one-loop effective potential.

To linear order in $\delta \phi$, we get  from (\ref{KG}):
\begin{eqnarray}
\Box \, \delta\phi +V''(\hat\phi)\delta\phi = 0 \,.
\end{eqnarray}
If the mass of the $\delta\phi$ field 
\begin{eqnarray}
m^2(\hat\phi)=V''(\hat\phi)
\end{eqnarray}
is constant, then the fluctutations can be canonically quantized, so that 
\begin{eqnarray}
\delta\phi (t,{\bf x})=\int \frac{\text{d}^3{\bf k}}{(2\pi)^{3/2}\sqrt{2\omega}}\, \left(a_{{\bf k}}\, e^{i\, ({\bf k}\,{\bf x}-\omega t)} + a^{\dag}_{{\bf k}}e^{-(i\, {\bf k}\,{\bf x}-\omega t)}\right) ,
\nonumber \\\label{fielddecomposition}
\end{eqnarray}
with
\begin{eqnarray}
\omega^2=k^2+m^2(\hat \phi) , \label{freq}
\end{eqnarray}
and $k^2={\bf k}\cdot{\bf k}$, 
where the creation and annihilation operators satisfy the standard commutation relations:
\begin{eqnarray}
[a_{{\bf p}},a^{\dag}_{{\bf q}}]=\delta^{(3)}({\bf p}-{\bf q})\, ,
\end{eqnarray}
and the annihilation operator defines the vacuum state:
\begin{eqnarray}
a_{{\bf p}}\vert 0\rangle=0, \;\;\forall {\bf p} .
\end{eqnarray}
Thus, using the expansion in (\ref{fielddecomposition}) we can compute
\begin{eqnarray}
\langle 0\vert \delta\phi^2\vert 0\rangle=\frac{1}{4\pi^2}\int_0^\infty dk \;\frac{k^2}{\sqrt{k^2+m^2(\hat\phi)}}, 
\end{eqnarray}
which is divergent, and can be regularized by using, for instance, dimensional regularization or a 
three-momentum cutoff. Thus, considering a cutoff $\Lambda$, we obtain for the second term in 
the effective potential in (\ref{flucteq}) the regularized expression:
\begin{eqnarray}
\frac{1}{2}V'''(\hat \phi)\langle 0\vert \delta\phi^2\vert 0\rangle_{reg}&=&
\frac{1}{8\pi^2}\frac{dm^2(\hat \phi)}{d\hat\phi}\int_0^\Lambda dk \;\frac{k^2}{\sqrt{k^2+m^2(\hat\phi)}}\nonumber \\
&=&\frac{dV_1(\hat\phi)}{d\hat \phi}, 
\end{eqnarray}
with
\begin{eqnarray}
V_{1}=\frac{1}{4\pi^2}\int_0^\Lambda dk \; k^2\sqrt{k^2+m^2(\hat\phi)}
\label{V1f}
\end{eqnarray}
being the one-loop contribution to the effective potential. Thus, we can rewrite (\ref{flucteq}) as follows
\begin{eqnarray}
\Box \, \hat\phi +V_{eff}'(\hat\phi) = 0 \, ,
\label{flucteqVeff}
\end{eqnarray}
where the effective potential reads 
\begin{eqnarray}
V_{eff}(\hat\phi)=V(\hat\phi)+ V_{1}(\hat\phi) .
\end{eqnarray}
The cutoff-regularized integral in (\ref{V1f}) can be calculated 
exactly as
\begin{eqnarray}
V_1(\hat\phi)&=&\frac{1}{32\pi^2}\left(\Lambda(2\Lambda^2+m^2(\hat\phi))\sqrt{\Lambda^2+m^2(\hat\phi)})\right.\nonumber \\
&+&\left. m^4(\hat\phi)\ln\left(\frac{m(\hat\phi)}{\Lambda+\sqrt{\Lambda^2+m^2(\hat\phi)}}\right)\right) .
\label{cutoffreg}
\end{eqnarray}
Therefore, in $\Lambda\rightarrow \infty$ limit we obtain
\begin{eqnarray}
V_1(\hat\phi)&=&{\frac {{\Lambda}^{4}}{16{\pi }^{2}}}+{\frac {{m}^{2}(\hat\phi){\Lambda}^{2}}{16{
\pi }^{2}}}-\frac{m^4(\hat\phi)}{64\pi^2}\ln\left(\frac{\Lambda^2}{\mu^2}\right)\nonumber \\
&+&
\frac{m^4(\hat\phi)}{64\pi^2}\ln\left(\frac{m^2(\hat\phi)}{\mu^2}\right)+\frac{m^4(\hat\phi)}{64\pi^2}\left(\frac{1}{2}-2\ln (2)\right)\nonumber\\
&+&\Od(\Lambda^{-2}) ,
\end{eqnarray}
where for convenience we have introduced an arbitrary scale $\mu$, which plays no 
role at this stage. 
We then have three types of divergences: quartic, quadratic and logarithmic, which can be eliminated
by including appropriate counter-terms. Thus, let us consider that the tree-level potential
is of the Higgs form, i.e.,
\begin{eqnarray}
V(\hat\phi)=V_0+\frac{1}{2}M^2\hat\phi^2+\frac{\lambda}{4}\hat\phi^4 ,
\end{eqnarray}
where $V_0$ is a cosmological constant contribution. 
Hence
\begin{eqnarray}
m^2(\hat\phi)=M^2+3\lambda \hat\phi^2 .
\end{eqnarray}
Thus, the divergent contributions read
\begin{eqnarray}
V_\infty(\hat\phi)&=&
\frac{{\Lambda}^{4}}{16{\pi }^{2}}+\frac {M^2{\Lambda}^{2}}{16{\pi}^{2}}-
\frac{M^4}{64\pi^2}\ln\left(\frac{\Lambda^2}{\mu^2}\right)\nonumber \\
&+&
\left(\frac{3\lambda\Lambda^2}{16\pi^2}-\frac{6\lambda M^2}{64\pi^2}\ln\left(\frac{\Lambda^2}{\mu^2}\right)\right)\hat\phi^2\nonumber \\
&-&\frac{9\lambda^2}{64\pi^2}\ln\left(\frac{\Lambda^2}{\mu^2}\right)\hat\phi^4 ,
\end{eqnarray}
which are either constant, quadratic or quartic in $\hat\phi$. It is therefore possible 
to absorbe all the divergences by including in the effective potential these  counter-terms as follows
\begin{eqnarray}
V_{eff}(\hat\phi)=V(\hat\phi)+ V_{1}(\hat\phi)+\Delta_{cc}+\frac{1}{2}\Delta_M\hat\phi^2+\frac{1}{4}\Delta_\lambda\hat\phi^4 , \nonumber \\ \label{count}
\end{eqnarray}
which is equivalent to a redefinition of the tree-level parameters. 
In the so-called minimal subtraction $\overline{MS}$ schemes,  the three counter-terms are chosen 
such that they exactly cancel the divergent contributions $V_{\infty}$ at a given $\mu$-scale,  so that 
for the renormalized effective potential we finally get
\begin{eqnarray}
V_{eff}(\hat\phi)&=&V_0+\frac{1}{2}M^2\hat\phi^2+\frac{\lambda}{4}\hat\phi^4 \label{Vren}\\
&+&
\frac{m^4(\hat\phi)}{64\pi^2}\left(\ln\left(\frac{m^2(\hat\phi)}{\mu^2}\right)+C\right),\nonumber
\end{eqnarray}
where $C$ is a constant which depends on the renormalization 
scheme. In our case $C=1/2-2\ln (2)$, but notice that this constant can be absorbed in a
$\mu$ redefinition.

Notice that the parameters $M$ and $\lambda$ are defined at the $\mu$-scale,  so that 
a given variation in $\mu$ should be compensated by a change in their values  in such 
a way that the renormalized effective potential remains unchanged. Hence, $M$ and $\lambda$  should 
depend on the renormalization scale $\mu$ according to (to leading order in 
$\lambda$):
 \begin{eqnarray}
\beta(\lambda)&\equiv&\frac{d\lambda}{d(\log \mu)}=\frac{18\lambda^2}{(4\pi)^2} ,\nonumber\\
\gamma_M(\lambda)&\equiv&\frac{d\log M^2}{d(\log \mu)}=\frac{6\lambda}{(4\pi)^2} .
\label{RGEf}
\end{eqnarray}
Notice that the minimum of the effective potential should be independent of the $\mu$ scale.
Also a change in the renormalization scheme is equivalent to a reparametrization 
of the mass and coupling constant. 

Equivalent results can be obtained by means of dimensional regularization. In that case,
the quadratic and quartic divergences are absent, but the renormalized $V_{eff}$ in the $\overline{MS}$
scheme 
agrees with  (\ref{Vren}) adopting $C=-3/2$.

For the loop expansion to make sense, the one-loop contribution $V_1$ should be small
as compared to the tree-level potential $V$.

\section{Equations in perturbed Robertson-Walker backgrounds}

We will extend the previous calculation to a flat Robertson-Walker background including scalar 
perturbations. We will work in the longitudinal gauge and the metric tensor takes 
the form 
\begin{eqnarray}
\text{d}s^2 = a^2(\eta) \left\{ \left[1 + 2 \Phi(\eta,{\bf x})\right]\, \text{d}\eta^2 - \left[1 - 2\Psi(\eta,{\bf x})\right]\,\text{d}{\bf x}^2 \right\}\,,\nonumber\\
\end{eqnarray} 
where $\eta$ is the conformal time, and $a$ the scale factor. Here, $\Phi$ and $\Psi$ are the perturbation potentials of the metric.

The action for the scalar field with potential $V(\phi)$ in curved space-time reads  as follows
\begin{eqnarray}
S[\phi]=\int d^4x \sqrt{g} \left(\frac{1}{2}g^{\mu\nu}\partial_\mu\phi\partial_\nu\phi-V(\phi)\right), 
\end{eqnarray}
and the corrresponding equations of motion
\begin{eqnarray}
\Box \, \phi +V'(\phi) = 0 \,,
\label{KGc}
\end{eqnarray}
up to first order in metric perturbations, can be written as
\begin{eqnarray}
\phi''&+&(2{\cal H}-\Phi'-3\Psi')\phi'- (1+2(\Phi+\Psi))\nabla^2\phi\nonumber \\
&-&\vec\nabla\phi\cdot\vec \nabla(\Phi-\Psi)
+a^2(1+2\Phi)V'(\phi)=0\,. \label{pert}
\end{eqnarray}

Here we have introduced ${\cal H}={a'}/a$. As in the flat space-time case presented above, 
the field $\phi$ can be decomposed
into a classical solution $\hat\phi(\eta ,{\bf x})$, which can be 
inhomogeneous because of the presence of metric perturbations, and a quantum fluctuation $\delta \phi(\eta,{\bf x})$, i.e.,
\begin{eqnarray}
\phi(\eta,{\bf x})=\hat \phi(\eta,{\bf x})+\delta \phi(\eta,{\bf x}). 
\end{eqnarray}
The effect of quantum fluctuations on the classical solutions can be taken into
account again by expanding the potential around the classical solution. Thus, the effective equation of motion
for the classical field reads
\begin{eqnarray}
\hat \phi''&+&(2{\cal H}-\Phi'-3\Psi')\hat\phi'- (1+2(\Phi+\Psi))\nabla^2\hat\phi\nonumber \\
&-&\vec\nabla\hat\phi\cdot\vec \nabla(\Phi-\Psi) \label{hat} \\
&+&a^2(1+2\Phi)\left(V'(\hat\phi)+\frac{1}{2}V'''(\hat\phi)\langle 0\vert\delta\phi^2\vert 0 \rangle\right)=0\,, \nonumber 
\end{eqnarray}
where again we have used $\langle 0\vert\delta\phi\vert 0 \rangle=0$ and neglected cubic and higher-order terms. 
Thus, once more, from the last term we can define the effective potential as
\begin{eqnarray}
V_{eff}(\hat\phi)=V(\hat\phi)+ V_{1}(\hat\phi), 
\end{eqnarray}
with 
\begin{eqnarray}
V_{1}'(\hat\phi)=\frac{1}{2}V'''(\hat\phi)\langle 0\vert\delta\phi^2\vert 0 \rangle .
\end{eqnarray}
The equation for the fluctuation field $\delta\phi$ can be obtained by linearizing (\ref{pert}), and reads
\begin{eqnarray}
&&\delta \phi''+(2{\cal H}-\Phi'-3\Psi')\delta\phi'- (1+2(\Phi+\Psi))\nabla^2\delta\phi\nonumber \\
&-&\vec\nabla\delta\phi\cdot\vec \nabla(\Phi-\Psi) +a^2(1+2\Phi)V''(\hat\phi)\delta\phi=0\,. \label{delta}
\end{eqnarray}
where the mass of the $\delta\phi$ field is again given by
\begin{eqnarray}
m^2(\hat\phi)=V''(\hat\phi), 
\end{eqnarray}
which can be considered as constant since the classical field solution 
 at tree level is just a constant field satisfying $V'(\hat\phi)=0$.

 In order to obtain the one-loop contribution to the effective potential $V_1$, we need to quantize the 
fluctuation field. Because of the presence of the metric perturbations in (\ref{delta}), 
the problem is now much more involved than in flat space-time, and hence,
the solutions cannot be obtained exactly. 
However, as we present below, it is possible
to obtain a perturbative expansion of the solutions in metric perturbations. Moreover, 
we also show that 
in the case in which the mode frequency  is larger than the typical frequency
of the temporal or spatial variations of the background metric, i.e. if $\omega^2\gg {\cal H}^2$
and $\omega^2\gg \{\nabla^2\Phi,\;\nabla^2\Psi\}$, it is possible to consider the standard adiabatic approximation in 
order to quantize the fluctuations. Notice that for the Standard Model Higgs field,  
the adiabatic approximation can be perfectly adopted during the whole radiation and 
matter eras until present, and
for all  cosmological and astrophysical scales of interest. 

\section{Quantization of the field fluctuations}
We will apply the canonical quantization procedure to the field perturbations $\delta \phi$. Thus following the approach in \cite{Albareti}, 
we need to obtain a complete set of  solutions for (\ref{delta}), 
which are orthonormal  
with respect to the curved space-time version of the standard scalar product
\cite{Birrell}:
\begin{eqnarray}
&&(\delta\phi_p,\delta\phi_q)= \\
&-&i\int_\Sigma \left[\delta\phi_p(x)\partial_\mu \delta\phi_q^*(x)
-(\partial_\mu \delta\phi_p(x))\delta\phi_q^*(x)\right]\sqrt{g_\Sigma}
\text{d}\Sigma^\mu , \nonumber 
\label{scalar}
\end{eqnarray}
where $\text{d}\Sigma^\mu=n^\mu \text{d}\Sigma$,  with $n^\mu$ being a unit timelike vector
directed to the future and orthogonal to the $\eta=\text{const.}$ hypersurface $\Sigma$, i.e.
\begin{eqnarray}
\text{d}\Sigma^\mu=\text{d}^3{\bf x} \left(\frac{1-\Phi}{a},0,0,0\right), 
\end{eqnarray}
and to first order in metric perturbations:
\begin{eqnarray}
\sqrt{g_\Sigma}=a^3(1-3\Psi) .
\end{eqnarray}
Defined in this way, the scalar product does not depend on the choice of spatial
hypersurface $\Sigma$. 

 Thus, we should have
\begin{eqnarray}
(\delta\phi_p,\delta\phi_q)=\delta^{(3)}({\bf p}-{\bf q}) ,
\label{normalization}
\end{eqnarray}
so that when quantizing
\begin{eqnarray}
\delta\phi(\eta,{\bf x})=\int \text{d}^3{\bf k} \left( a_{{\bf k}}\delta\phi_k(\eta,{\bf x})+a^{\dag}_{{\bf k}}\delta\phi_k^*(\eta,{\bf x})\right) ,
\end{eqnarray}
the corresponding creation and annihilation operators would satisfy the
standard commutation relations:
\begin{eqnarray}
[a_{{\bf p}},a^{\dag}_{{\bf q}}]=\delta^{(3)}({\bf p}-{\bf q})\,.
\end{eqnarray}

The orthonormal set can be written using the WKB ansatz:
\begin{eqnarray}
\delta\phi_k(\eta,{\bf x})=f_k(\eta,{\bf x}) \,e^{i\theta_k(\eta,{\bf x})} ,
\label{wkb}
\end{eqnarray}
where $f_k(\eta,{\bf x})$ evolves slowly with $\eta$ and ${\bf x}$, whereas
$\theta_k(\eta,{\bf x})$ is rapidly evolving. As commented above,  
such an ansatz is expected to work when the Compton wavelength of the field 
is much smaller than the typical cosmological scales.

Substituting in (\ref{delta}), we get to the leading adiabatic order $\Od(\theta^2)$:
\begin{eqnarray}
-\theta'^2_k+(\vec\nabla\theta_k)^2(1+2(\Phi+\Psi))+m^2a^2(1+2\Phi)=0\,. \nonumber \\ \label{leading}
\end{eqnarray}
The next to leading term $\Od(\theta)$ of  (\ref{delta})  reads
\begin{eqnarray}
2f'_k\theta'_k&+&f_k\theta''_k+f_k\theta'_k(2{\cal H}-\Phi'-3\Psi')\nonumber \\
&-&2\vec\nabla f_k\cdot\vec\nabla\theta_k-f_k\nabla^2\theta_k=0 . \label{next}
\end{eqnarray}
These equations can now be solved perturbatively in metric perturbations.
 Thus, to the lowest order
(\ref{delta}) reads:
\begin{eqnarray}
\delta {\phi^{(0)}}''+2{\cal H}\delta{\phi^{(0)}}'- \nabla^2\delta\phi^{(0)}
+a^2m^2(\hat\phi)\delta\phi^{(0)}=0\, ,
\end{eqnarray}
where $a^2m^2(\hat\phi)$ only depends on time. It is therefore possible to find solutions by Fourier transformation
in the spatial coordinates. Thus, 
the positive frequency solution  with momentum ${\bf k}$
can be written as
\begin{eqnarray}
\delta\phi^{(0)}_k(\eta,{\bf x})=F_k(\eta) e^{i{\bf k}\cdot{\bf x}-i\int^\eta\omega(\eta')\text{d}\eta'} ,
\end{eqnarray}
with
\begin{eqnarray}
\omega^2=k^2+m^2a^2 \label{omega} ,
\end{eqnarray}
and 
\begin{eqnarray}
F_k(\eta)=\frac{1}{a(2\pi)^{3/2}\sqrt{2\omega}} ,
\label{Fsol}
\end{eqnarray}
in order to have the correct normalization given in (\ref{normalization}).

Hence, the expansion of (\ref{wkb}) in metric perturbations can be performed by expanding the 
amplitude and  phase as follows
\begin{eqnarray}
f_k(\eta,{\bf x})&=&F_k(\eta)+\delta f_k(\eta,{\bf x})\nonumber \\
\theta_k(\eta,{\bf x})&=&-\int^\eta\omega(\eta')\,\text{d}\eta'+{\bf k}\cdot {\bf x}+\delta\theta_k(\eta,{\bf x}) . \label{at}
\end{eqnarray}
Substituting (\ref{at}) in (\ref{leading}), we obtain  (\ref{omega}) to the lowest order as expected, 
and to the first order in metric perturbations we obtain
\begin{eqnarray}
2\omega\,\delta\theta'_k+2k^2(\Phi+\Psi)+2{\bf k}\cdot \nabla\delta\theta_k
+2m^2a^2\Phi=0\, . \nonumber \\ \label{leadingpert}
\end{eqnarray}
On the other hand, substituting in (\ref{next}) we recover (\ref{Fsol}) to the lowest order,
whereas to first order in metric perturbations we get:
\begin{eqnarray}
&-&2\omega\delta f'_k+2F'_k\delta\theta'_k+F_k\delta\theta''_k-\omega'\delta f_k\nonumber \\
&-&2\omega{\cal H}\delta f_k+\omega F_k\Phi'+3\omega F_k\Psi'+2 F_k {\cal H}\delta\theta'_k\nonumber \\
&-&2{\bf k}\cdot \nabla\delta f_k-F_k\nabla^2\delta\theta_k 
-F_k {\bf k}\cdot \nabla(\Phi-\Psi)=
0\,. \label{nonleadingpert}
\end{eqnarray}

\section{Perturbative solutions}
We are now in the position to figure out how to find solutions to the perturbative equations (\ref{leadingpert})
and (\ref{nonleadingpert}). With that purpose we will perform an additional Fourier transform.
Notice that $\delta\theta$, $\Phi$ and $\Psi$ are functions of $(\eta, {\bf x})$, but the 
coefficients in those equations are only functions of time, so that we can Fourier transform
with respect to the spatial coordinates. Hence (\ref{leadingpert}) can be rewritten as follows

\begin{eqnarray}
2&\omega&\,\left(\delta\theta'_k(\eta,{\bf p})+i\frac{{\bf k}\cdot{\bf p}}{\omega}\delta\theta_k(\eta,{\bf p})\right)\nonumber \\
&=&-2\omega^2\Phi(\eta,{\bf p})-2k^2\Psi(\eta,{\bf p}) ,
\label{deltagen}
\end{eqnarray}
where 
\begin{eqnarray}
\delta\theta_k(\eta,{\bf p})=\frac{1}{(2\pi)^{3/2}}\int \text{d}^3{\bf x} \,
\delta\theta_k(\eta,{\bf x})e^{i{\bf p}\cdot{\bf x}}, 
\end{eqnarray}
and similar definitions for $\Phi(\eta,{\bf p})$, $\Psi(\eta,{\bf p})$ and $\delta f_k(\eta,{\bf p})$. Hereafter, we will use ${\bf k}$
 to denote the quantum fluctuation wavevector, and ${\bf p}$ for the wavevector of metric perturbations. Notice that 
strictly speaking, in the
presence of perturbations,  ${\bf k}$ cannot be understood as the comoving three-momentum of the quantum fluctuations,  
but simply as a way to label the different mode solutions.
Defining:
\begin{eqnarray}
\alpha_k(\eta,{\bf p})&=&\frac{{\bf k}\cdot{\bf p}}{\omega}\nonumber \\
\beta_k(\eta,{\bf p})&=&\int_0^\eta \alpha_k(\eta',{\bf p}) d\eta' \nonumber \\
G_k(\eta,{\bf p})&=&-\omega\Phi(\eta,{\bf p})-\frac{k^2}{\omega}\Psi(\eta,{\bf p}), 
\end{eqnarray}
we can rewrite (\ref{deltagen}) as
\begin{eqnarray}
\delta\theta'_k(\eta,{\bf p})+i\alpha_k(\eta,{\bf p})\delta\theta_k(\eta,{\bf p})=G_k(\eta,{\bf p}), 
\end{eqnarray}
whose solution reads
\begin{eqnarray}
\delta\theta_k(\eta,{\bf p})=e^{-i\beta_k(\eta,{\bf p})}\int_{0}^\eta e^{i\beta_k(\eta',{\bf p})} \, G_k(\eta',{\bf p})
d\eta' .
\label{dtheta}
\end{eqnarray}
Here we have chosen the integration limits so that
the perturbed phases match the unperturbed ones at $\eta=0$.
On the other hand, equation (\ref{nonleadingpert}) can be rewritten after Fourier transformation as
\begin{eqnarray}
\frac{2\sqrt{\omega}}{a}e^{-i\beta_k(\eta,{\bf p})}\left(e^{i\beta_k(\eta,{\bf p})} a\sqrt{\omega}\,\delta f_k(\eta,{\bf p})\right)' =F_k(\eta)H_k(\eta,{\bf p}) \nonumber \\
\end{eqnarray}
where
\begin{eqnarray}
&H_k(\eta,{\bf p})&=\omega\left(-i\frac{\alpha_k(\eta,{\bf p})}{\omega}\delta\theta_k(\eta,{\bf p})+ 
\Psi(\eta,{\bf p})\left(3-\frac{k^2}{\omega^2}\right)\right)'\nonumber \\
&+&p^2\delta\theta_k(\eta,{\bf p})-i\alpha_k(\eta,{\bf p})(\Phi(\eta,{\bf p})-\Psi(\eta,{\bf p}))\omega;
\end{eqnarray}
so that its solution is given by
\begin{eqnarray}
\delta f_k(\eta,{\bf p})=F_k(\eta)P_k(\eta,{\bf p}),
\label{df}
\end{eqnarray}
with
\begin{eqnarray}
P_k(\eta,{\bf p})=e^{-i\beta_k(\eta,{\bf p})}\left(\int_0^\eta e^{i\beta_k(\eta',{\bf p})}\frac{H_k(\eta',{\bf p})}{2\omega}d\eta'+D_k({\bf p})\right)\nonumber \\
\end{eqnarray}
The integration constant $D_k({\bf p})$ in the last expression is fixed by the normalization
condition (\ref{normalization}). In the particular cases we will consider below in which the gravitational
potentials are time-independent, it is given by:
\begin{eqnarray}
D_k({\bf p})= \frac{1}{2}\left(3-\frac{k^2}{\omega_0^2}\right)\Psi({\bf p})
\end{eqnarray}
with $\omega_0=\omega(\eta=0)$.
 
\section{Effective potential}
Once we have the expressions for the mode solutions to the 
perturbative equations,  we can proceed to calculate the one-loop contribution
to the effective potential. Let us first calculate $\langle 0\vert\delta\phi^2(\eta, {\bf x})\vert 0 \rangle$ to first order in metric perturbations. Notice that because of the inhomogeneity 
of the background, this quantity depends on $(\eta, {\bf x})$ as follows
\begin{eqnarray}
&&\langle 0\vert\delta\phi^2(\eta, {\bf x})\vert 0 \rangle=\int \text{d}^3{\bf k}\,F_k^2(\eta)
\nonumber  \\
&+&2\int \text{d}^3{\bf k}\, F_k(\eta)\,({\mbox Re}\,\delta  f_k(\eta, {\bf x})-F_k(\eta){\mbox Im }\,\delta\theta_k(\eta, {\bf x}))
\nonumber \\&=&\Delta_{h}(\eta)+\Delta_{i}(\eta, {\bf x}) .
\end{eqnarray}
The regularized homogeneous contribution given by the first term in the 
integral reads
\begin{eqnarray}
\Delta_{h}(\eta)=\frac{1}{4\pi^2a^2}\int_0^\Lambda dk  \frac{k^2}{\sqrt{k^2+m^2a^2}}, 
\end{eqnarray}
and is analogous to the Minkowskian result, except for the scale-factor dependence.

The inhomogeneous component $\Delta_i$, can be computed more easily in momentum 
space. Thus, since the metric potentials only depend on $p$, but not on
the momentum direction,  then, the only
angular dependence enters in $\alpha(\eta,{\bf p})=(kp\hat x/\omega)$ with $\hat x=\cos \theta$, where 
we have taken the $k_z$ direction along  ${\bf p}$. Hence, we can write
\begin{eqnarray}
\Delta_{i}(\eta, {\bf p})&=&\frac{1}{4\pi^2a^2}\int_0^\Lambda dk  \frac{k^2}{\sqrt{k^2+m^2a^2}}\int_{-1}^1 d\hat x\, P_k(\eta,p,\hat x), \nonumber \\
\end{eqnarray}
which is actually isotropic and only depends on $p$.

We can now calculate the one-loop contribution to the effective potential:
\begin{eqnarray}
V_{1}&=&\frac{1}{2}\int dm^2 \langle 0\vert \delta\phi^2(\eta,{\bf x})\vert 0\rangle_{reg}\nonumber\\
&=& V_1^{h}(\eta)+V_1^i(\eta, {\bf x}) .
\end{eqnarray}
Thus, the homogeneous contribution reads
\begin{eqnarray}
V_1^{h}(\eta)=
\frac{1}{4\pi^2 a^4}\int_0^\Lambda dk \; k^2\sqrt{k^2+a^2m^2} ,
\label{RWint}
\end{eqnarray}
whereas the inhomogeneous one is given in Fourier space by
\begin{eqnarray}
&&V_1^i(\eta, {\bf p}) \label{V1}\\
&=&\frac{1}{8\pi^2a^2}\int dm^2\int_0^\Lambda dk  \frac{k^2}{\sqrt{k^2+m^2a^2}}\int_{-1}^1 d\hat x\, P_k(\eta,p,\hat x). \nonumber
\end{eqnarray}
It can be seen that after integrating in $\hat x$ the phase perturbation
$\delta\theta$ does not contribute to the final result.
\section{Renormalization}
Let us now discuss the renormalization procedure in a non-trivial
background geometry. First, we consider the homogeneous contribution $V_1^h$.
The momentum integral in (\ref{RWint}) can be done exactly as in (\ref{cutoffreg}), and 
 in the $\Lambda\rightarrow \infty$ limit we obtain
\begin{eqnarray}
V_1^h(\hat\phi)&=&{\frac {{\Lambda}^{4}}{16{\pi }^{2}a^4}}+{\frac {{m}^{2}(\hat\phi){\Lambda}^{2}}{16{
\pi }^{2}a^2}}-\frac{m^4(\hat\phi)}{64\pi^2}\ln\left(\frac{\Lambda^2}{\mu^2}\right)\nonumber \\
&+&
\frac{m^4(\hat\phi)}{64\pi^2}\ln\left(\frac{m^2(\hat\phi)a^2}{\mu^2}\right)+\frac{m^4(\hat\phi)}{64\pi^2}\left(\frac{1}{2}-2\ln (2)\right)\nonumber\\
&+&\Od(\Lambda^{-2}), 
\end{eqnarray}
where $\mu$ is a comoving renormalization scale. 
Notice that the  quartic and quadratic divergences depend on the scale 
factor, whereas the logarithmic divergence is constant. Notice, however, that the finite terms
contain constant and $\ln a$ terms.
 As a matter of fact, $V_1^h$ 
can be interpreted as the vacuum energy of the 
scalar field fluctuations in a Robertson-Walker background.
Thus, in order to eliminate the divergences, it would be necessary to add conserved counterterms
to the energy-momentum tensor.  Notice also, that in dimensional regularization,
the absence of quadratic and quartic divergences implies that the required counterterms are
scale-factor independent.

Thus, following the minimal subtraction scheme discussed before, we are 
left with the renormalized homogeneous contribution to the effective potential, i.e.
\begin{eqnarray}
V_{eff}^h(\hat\phi)&=&V_0+\frac{1}{2}M^2\hat\phi^2+\frac{\lambda}{4}\hat\phi^4 \label{Vren}\\
&+&
\frac{m^4(\hat\phi)}{64\pi^2}\left(\ln\left(\frac{m^2(\hat\phi)}{\mu_{ph}^2}\right)+C\right) ,\nonumber
\end{eqnarray}
where the physical renormalization scale is given by $\mu_{ph}=\mu/a$ and,  
as commented above, $C$ is a constant which depends on the 
regularization method and the renormalization scheme. Here,  the physical mass $M$ and coupling constant $\lambda$ are defined
at a given physical scale $\mu_{ph}$.
In addition, the renormalized energy-momentum tensor is independent of the renormalization
scale $\mu_{ph}$. This is again guaranteed by the $\mu_{ph}$ dependence of the mass and coupling 
constant given by (\ref{RGEf}).

Let us now consider the inhomogeneous contribution. In order to obtain explicit expressions, we adopt the  case 
of a matter dominated universe in which the gravitational potentials $\Phi({\bf p})$ and 
$\Psi({\bf p})$ do not depend on time.  In this case,
 after a  lengthy but straightforward
calculation we obtain
\begin{eqnarray}
&&V_{eff}^i(\hat\phi)=\frac{m^{2}(\hat\phi){\Lambda}^2}{16{
\pi}^{2}a^2}{\cal F}\left[\frac{\sin (p\eta)}{p\eta}(\Phi+3\Psi)-(\Phi+\Psi)\right] \nonumber \\
&+&\frac{m^4(\hat\phi)}{64\pi^2}\ln\left(\frac{m^2(\hat\phi)a^2}{\Lambda^2}\right){\cal R}(\eta,{\bf x}),  
+\Od(\Lambda^{-2})\label{matter}
\end{eqnarray}
with 
\begin{eqnarray}
&&{\cal R}(\eta,{\bf x})\nonumber \\
&=&{\cal F}\left[(\Phi({\bf p})-\Psi({\bf p}))\left( 1-\frac{1}{5}\left(\cos (p\eta)+4\frac{\sin (p\eta)}{p\eta}\right)\right)\right]. 
\nonumber \\
\end{eqnarray}
Here, ${\cal F}$ denotes the  Fourier transform.  

Notice 
that no quartic divergences are present in the inhomogeneous component in agreement with \cite{Huanginho}. The quadratic and 
logarithmic divergences are local 
in  $\hat \phi$ and contain the same powers as  those in flat space-time, so that they can be eliminated
by adding the same kind of local counterterms in $\hat  \phi$. The only difference 
comes from the fact that the coefficients depend now on the space-time position.
This implies that the required counterterms have also space-time dependent 
coefficients. Notice that this renormalization procedure is consistent, 
 since the renormalized energy-momentum tensor obtained in this way is still covariantly conserved. 
After removing the divergences we get
\begin{eqnarray}
V_{eff}^i(\hat\phi)&=&\frac{m^4(\hat\phi)}{64\pi^2}\ln\left(\frac{m^2(\hat\phi)}{\mu_{ph}^2}\right){\cal R}(\eta,{\bf x}) 
\nonumber 
\end{eqnarray}
up to finite renormalization constans, which can be absorbed in the renormalization scale $\mu_{ph}$. 
 As we are working at the leading order in the 
adiabatic expansion, the divergences do not contain derivatives of the metric tensor. Subleading contributions are expected to include divergences with two and four metric derivatives, 
which correspond to terms linear and quadratic in curvatures \cite{ParkerFulling}.  In the cosmological contexts
in which we are interested, i.e. at late times, those terms can be safely neglected.

Finally, we can write for the total effective potential $V_{eff}$ the following expression
\begin{eqnarray}
V_{eff}(\hat\phi)&=&V_{eff}^h(\hat\phi)+V_{eff}^i(\hat\phi)=
V_0+\frac{1}{2}M^2\hat\phi^2+\frac{\lambda}{4}\hat\phi^4 \nonumber\\
&+&
\frac{m^4(\hat\phi)}{64\pi^2}\ln\left(\frac{m^2(\hat\phi)}{\mu_{ph}^2}\right)
(1+{\cal R}(\eta,{\bf x})).\label{Vren}
\end{eqnarray}
In the case in which we use a constant $\mu_{ph}$ scale, the homogeneous potential behaves as 
a cosmological constant. This means that although the calculation of the metric perturbation correction
has been done in the longitudinal gauge, the Stewart-Walker lemma guarantees that 
the effective potential is gauge invariant. Notice that the effective potential
given in (\ref{Vren}) only includes
Higgs field fluctuations. Due to the couplings to quarks and gauge bosons, 
we would expect additional contributions coming mainly from W$^{\pm}$ and Z gauge bosons, and top quarks
\cite{ColemanWeinberg}.

In the case of super-Hubble fluctuations $p\eta\ll 1$, the effective potential reads 
\begin{eqnarray}
V_{eff}(\hat\phi)&=&V_{eff}^h(\hat\phi)+V_{eff}^i(\hat\phi)=
V_0+\frac{1}{2}M^2\hat\phi^2+\frac{\lambda}{4}\hat\phi^4 \nonumber\\
&+&
\frac{m^4(\hat\phi)}{64\pi^2}\ln\left(\frac{m^2(\hat\phi)}{\mu_{ph}^2}\right)
\left[1+\Od(p^2\eta^2)\right],
\end{eqnarray}
i.e. there is no inhomogeneous contribution to leading order in the super-Hubble limit. This result holds also in the radiation era. Notice that this is consitent
with the fact that for homogeneous perturbations, i.e. $p\rightarrow 0$,
we recover the standard homogeneous expression for the effective potential.
 
The fact that the divergences are space-time dependent, because of the presence of the inhomogeneos background, 
implies that the running of the mass and coupling constant should also depend
on the space-time position. Thus, we get
\begin{eqnarray}
\beta(\lambda)&\equiv&\frac{d\lambda}{d(\log \mu_{ph})}=\frac{18\lambda^2}{(4\pi)^2}(1+{\cal R}(\eta,{\bf x})) ,\nonumber\\
\gamma_M(\lambda)&\equiv&\frac{d\log M^2}{d(\log \mu_{ph})}=\frac{6\lambda}{(4\pi)^2}(1+{\cal R}(\eta,{\bf x})) .
\label{RGE}
\end{eqnarray}
The space-time dependence of the beta functions required for the correct renormalization
in this case has been invoked also in the context of the local renormalization group  related to 
local Weyl rescaling of the metric when the coupling constants are also allowed to be space-time 
dependent \cite{osborn}.

In the case of a perturbed static universe, i.e. a perturbed Minkowski 
metric in which the gravitational potentials $\Phi(p)$ and 
$\Psi(p)$ do not depend on time, we get the same result as in the 
matter dominated universe given in (\ref{matter}), but with
\begin{eqnarray}
{\cal R}(\eta,{\bf x})={\cal F}\left[(\Phi({\bf p})-\Psi({\bf p}))\left( 1-\cos (p\eta)\right)\right], 
\end{eqnarray}
i.e. the quartic and quadratic divergences are insensitive to the universe expansion, whereas the logarithmic 
terms change. Also in this case, in the limit $p\rightarrow 0$, i.e. taking the curvature radius of the spatial sections to infinite, the 
contributions to  ${\cal R}$ vanish.  

\section{Higgs vacuum expectation value}
The minimum of the effective potential  
\begin{eqnarray}
{V_{eff}}'(\hat\phi_{vac})=0
\end{eqnarray}
will determine the vacuum expectation value
of the Higgs field. In our case, the inhomogeneous contribution will induce a space-time dependence
in $\hat\phi_{vac}$. We will compute the minimum perturbatively in ${\cal R}$. Thus, if 
we  assume
\begin{eqnarray}
\hat\phi_{vac}=\hat\phi_0+\Delta\hat\phi, 
\end{eqnarray}
 where $\hat\phi_0$ is the minimum of the potential in the absence of metric
perturbations, i.e.
\begin{eqnarray}
{V_{eff}^h}'(\hat\phi_0)=0 ,
\end{eqnarray}
then:
\begin{eqnarray}
\Delta\hat\phi=-\frac{{V_{eff}^i}'(\hat\phi_0)}{{V_{eff}^h}''(\hat\phi_0)}.
\end{eqnarray}
In the case of the Higgs potential, $M^2<0$,  the tree-level vacuum expectation value is $v^2=-M^2/\lambda$.
The corresponding tree-level Higgs mass is $V''(v)=m_H^2=-2M^2$.
Taking the renormalization scale $\mu_{ph}=m_H, $we find 
\begin{eqnarray}
\frac{\Delta\hat\phi}{\hat\phi_0}=-\frac{3\lambda}{32\pi^2}{\cal R} 
=-\frac{3}{256\pi^2}{\cal R}
\end{eqnarray}
where $\lambda=m_H^2/2v^2$. Here  we have made use of the fact that for the resonance observed at LHC 
with tree-level parameters $m_H\simeq 125$ GeV, and $v\simeq 250$ GeV, 
the Higgs self-coupling is given by $\lambda\simeq 1/8$. 

Notice that the perturbation ${\cal R}$ can be written in terms of the
so called gravitational slip \cite{slip} (corresponding to the Eddington post-Newtonian parameter
for local gravity tests \cite{Will}) 
\begin{eqnarray}
\varpi=1-\eta=\frac{\Phi-\Psi}{\Phi} ,
\end{eqnarray}
which is small but non-vanishing in standard cosmology due to the  contributions from
neutrino diffusion and second order perturbations. Contributions to the gravitational
slip also appear generically in modified gravity theories  \cite{Amendola}. 
On large scales $p\simeq 10^{-4}$ h/Mpc, the gravitational slip could
reach   $\varpi\simeq 10^{-3}$ \cite{Ballesteros,Saltas}. 
Hence, we can estimate ${\cal R}\simeq \Phi  \varpi$, and therefore 
\begin{eqnarray}
\frac{\Delta \phi}{\phi} \simeq  10^{-3}\Phi\varpi .
\label{variation}
\end{eqnarray}

 For typical values of the metric perturbations on cosmological scales ${\cal R}\simeq \Phi  \varpi \simeq 10^{-5} \varpi$, we expect fluctuations in the Higgs vacuum expectation value of the order 
\begin{eqnarray}
\frac{\Delta \phi}{\phi}\simeq  10^{-8}\varpi \simeq 10^{-11} .\label{cosmo}
\end{eqnarray}

In principle, the fluctuations in the Higgs VEV could lead to large fluctuations in the vacuum energy itself which could
conflict with observations.  However as we show below this is not the case.  Let us evaluate the effective potential (\ref{Vren}) at the minumum, i.e.
$V_{eff}(\hat\phi_{vac})$. In the loop expansion, the one-loop contribution $V_1$ has to be evaluated for the tree-level value of the field 
$\hat\phi_{vac}\simeq v$. Accordingly, taking into 
account that $\mu^2_{ph}=m_H^2=m^2(v)$,  the logarithm of the one-loop contribution vanishes. This 
implies that $V^i_{eff}(\hat\phi_{vac})=0$ at the one-loop level. 
On the other hand, by expanding the remaining terms around $\hat\phi_0$ and 
using the fact that ${V_{eff}^h}'(\hat\phi_0)=0$,  we see that, to first order
in metric perturbations, $V_{eff}(\hat\phi_{vac})=V(v)$, i.e. 
the renormalized potential is  just a constant at the potential minimum. 
Accordingly, taking the constant term of the tree-level potential as $V_0=M^4/(4\lambda)$, it can be easily seen that $V_{eff}(\hat\phi_{vac})=0$, i.e.
even though the Higgs VEV depends on the space-time position, the vacuum energy is just a constant that can be cancelled by the constant $V_0$ term
of the potential, as in flat space-time. 

On the other hand,  the Higgs mass $\bar m^2=V''_{eff}(\hat\phi_{vac})$ acquires space-time fluctuations which can be evaluated in a
straightforward way. Writing $\bar m^2=m_0^2+\Delta m^2$, with $m_0^2={V^h_{eff}}''(\hat\phi_0)$, we get
\begin{eqnarray}
\frac{\Delta m^2}{m_0^2}=\frac{21\lambda}{32\pi^2}{\cal R} 
\simeq \frac{21}{256\pi^2}{\cal R} .
\end{eqnarray}
Notice also that despite the fact that the  VEV is space-time dependent, the contribution of gradients or time 
derivatives are suppressed by $\Od(\nabla^2/m^2)$ with respect to the potential
contribution in the equations of motion (\ref{hat}), thus confirming the validity of the effective potential approach in this case.

\section{Discussion and conclusions}
The calculation presented in this work is based on the construction of an 
explicit set of orthonormal  adiabatic mode solutions of the perturbed Klein-Gordon equation
using comoving coordinates. This set of modes differs from that obtained
by using Riemann normal coordinates, which locally agree with the flat space-time
expressions and therefore eliminates locally the metric dependence.
This fact explains why the inhomogeneous contributions are not apparent to the lowest order
in the Schwinger-de Witt approach.  In that case, the inhomogeneous contributions
to the effective potential appear only at second order in metric derivatives.

Regarding the regularization procedure, the inhomogeneous contribution of
the effective potential 
was obtained by means of 
a three-momentum cutoff. It would be interesting to compare these results with
those obtained using dimensional regularization. However, the complicated 
$k$ dependence of the integrands in (\ref{V1}) prevented us from using 
standard dimensional regularization expressions in this case. 

It is also important to stress that quantum field theory does not predict the 
value of any physical (renormalized) parameter since they can only be obtained from
experiments. In this sense, the results of the work presented here simply suggest a space-time 
dependence of the Higgs VEV and of the particle masses. However, the actual
variation can only be obtained from observations. In this respect, current observational 
limits on the temporal variation of the  electron mass or on the proton to electron 
mass ratio \cite{Uzan} imposes stringent limits on the Higgs VEV variation. Thus,  
limits on cosmological scales from high-redshift quasar absortion spectra imply 
$\Delta \phi/\phi\lsim 10^{-6}$, which is compatible with our estimate given 
in (\ref{cosmo}). 
On the other hand, on local scales, although there are limits from molecular clouds 
spectra in the Milky Way, 
the most
stringent constraints come for  atomic clock experiments on Earth,  which 
set limits   $\Delta \phi/\phi\lsim 10^{-16}$ \cite{Uzan}. Taking into 
account that  $\Phi\simeq 10^{-9}$ on the surface of the Earth, using (\ref{variation}) we can  translate
 this limit into a limit on the Eddington parameter on Earth around $\varpi<10^{-4}$. Notice that existing limits on $\varpi$ are based on light deflection by 
the Sun or on Shapiro time delay of signals passing also close to the 
surface of the Sun.  Since the gravitational potential on the surface of the Sun is much larger, around 
$\Phi\simeq 10^{-6}$, atomic clocks experiments near the Sun  would allow to set more
stringent limits,  
which could reach $\varpi\lsim 10^{-7}$. Such limits could improve the existing ones from
Doppler tracking of the Cassini spacecraft \cite{Bertotti},  which are around 
$\varpi\lsim 2\times 10^{-5}$.
Better sensitivity  could be obtained in regions with even stronger gravitational
fields, such as for instance, the galactic center near the supermassive black hole. Observations of spectra
from objects in those regions could thus provide an independent probe of the electroweak symmetry breaking sector.

\section{Erratum}
After publication of the paper, a term missing in equation (42)
 corresponding to the next to leading order in the adiabatic approximation was identified. The 
corrected equation now reads
\begin{eqnarray}
f_k\theta''_k &+& 2f'_k\theta'_k +\left[2{\cal H}-\Phi'-3\,\Psi'\right] f_k\theta'_k \\
&-&f_k\nabla^2\theta_k - 2\boldsymbol{\nabla} f_k\cdot\boldsymbol{\nabla}\theta_k\,-f_k \,\boldsymbol{\nabla}\theta_k\cdot \boldsymbol{\nabla}\left[\Phi-\Psi\right]=\,0\,.\nonumber \label{next}
\end{eqnarray}

This implies that (49) should read
\begin{eqnarray}
&&F_k\, \delta\theta''_k+ 2F'_k\,\delta\theta'_k +2{\cal H}\,F_k \,\delta\theta'_k -F_k\,\nabla^2\delta\theta_k\nonumber \\ 
&-& 2\,\omega\,\delta f'_k-2\,{\bf k}\cdot \boldsymbol{\nabla}\delta f_k
-2\omega\,{\cal H}\,\delta f_k -\omega'\delta f_k \label{nonleadingpert} \\ 
&+& \omega\, F_k\,\Phi'+3\,\omega \,F_k\,\Psi'
- F_k \,{\bf k}\cdot \boldsymbol{\nabla}\left[\Phi-\Psi\right]\,=\,
0\,.\nonumber 
\end{eqnarray}

Then equation (56) now reads
\begin{eqnarray}
&H_k(\eta,{\bf p})&=\omega\left(-i\frac{\alpha_k(\eta,{\bf p})}{\omega}\delta\theta_k(\eta,{\bf p})+ 
\Psi(\eta,{\bf p})\left(3-\frac{k^2}{\omega^2}\right)\right)'\nonumber \\
&+&p^2\delta\theta_k(\eta,{\bf p})-i\alpha_k(\eta,{\bf p})(\Phi(\eta,{\bf p})-\Psi(\eta,{\bf p}))\omega
\end{eqnarray}
The new contribution implies that the cutoff expansion of the inhomogeneous contribution (68) is
\begin{eqnarray}
&&V_{eff}^i(\hat\phi)=\frac{m^{2}(\hat\phi){\Lambda}^2}{8{
\pi}^{2}a^2}{\cal F}\left[\frac{\sin (p\eta)}{p\eta}(\Phi+\Psi)-\Phi\right]   
+\Od(\Lambda^{-2})\nonumber \\\label{matter}
\end{eqnarray}
In other words, the logarithmic divergences exactly vanish. Accordingly, 
the analysis on the variations of the Higgs VEV 
based on the logarithmic contribution  does not apply.

\vspace*{0.1cm}

{\bf Acknowledgements}
We would like to thank S. Odintsov and E. Elizalde for useful comments.
This work has been supported by MICINN (Spain) project numbers FIS2011-23000, AYA2010-21231-C02-01,  
Consolider-Ingenio MULTIDARK CSD2009-00064
and Centro de Excelencia Severo Ochoa Programme under grant SEV-2012-0249.


\vspace{0.1cm}
\begin{thebibliography}{0}
\vspace{0.1cm}




\bibitem{Higgs1}
  G.~Aad {\it et al.}  [ATLAS Collaboration],
  Phys.\ Lett.\ B {\bf 716} (2012) 1
  [arXiv:1207.7214 [hep-ex]]
\bibitem{Higgs2}
S.~Chatrchyan {\it et al.}  [CMS Collaboration],
  Phys.\ Lett.\ B {\bf 716} (2012) 30
  [arXiv:1207.7235 [hep-ex]].
\bibitem{Higgsprop}
  G.~Belanger, B.~Dumont, U.~Ellwanger, J.~F.~Gunion and S.~Kraml,
  Phys.\ Rev.\ D {\bf 88} (2013) 075008
  [arXiv:1306.2941 [hep-ph]];
\bibitem{future} M.~V.~Marono [for the CMS Collaboration],
  arXiv:1409.1711 [hep-ex].
\bibitem{Mann}
  C.~Alvarez and R.~B.~Mann,
  Phys.\ Rev.\ D {\bf 54} (1996) 5954
  [gr-qc/9507040] and 
  Gen.\ Rel.\ Grav.\  {\bf 29} (1997) 245
  [gr-qc/9605039].

\bibitem{Maggiore}M.~Maggiore,
  Phys.\ Rev.\ D {\bf 83} (2011) 063514
  [arXiv:1004.1782 [astro-ph.CO]].
\bibitem{Sola}J.~Sol\'a,
  J.\ Phys.\ Conf.\ Ser.\  {\bf 453} (2013) 012015
  [arXiv:1306.1527 [gr-qc]];
J.~Sol\'a,
  Int.\ J.\ Mod.\ Phys.\ A {\bf 29} (2014) 2,  1444016
  [arXiv:1408.4427 [gr-qc]].

\bibitem{SdW}
  J.~S.~Schwinger,
  Phys.\ Rev.\  {\bf 82} (1951) 914;
 B.~S.~DeWitt,
  Phys.\ Rept.\  {\bf 19} (1975) 295;
P.~C.~W.~Davies, S.~A.~Fulling, S.~M.~Christensen and T.~S.~Bunch,
  Annals Phys.\  {\bf 109} (1977) 108;
T.~S.~Bunch and P.~C.~W.~Davies,
  J.\ Phys.\ A {\bf 11} (1978) 1315; 
I.L. Buchbinder, S.D. Odintsov, I.L. Shapiro, {\it  Effective action in quantum gravity}, Tomsk Pedagogical Inst. (1992). 
\bibitem{normal}
F.~Sobreira, B.~J.~Ribeiro and I.~L.~Shapiro,
  Phys.\ Lett.\ B {\bf 705} (2011) 273
  [arXiv:1107.2262 [gr-qc]]
\bibitem{Shapiro}
 M.~Asorey, P.~M.~Lavrov, B.~J.~Ribeiro and I.~L.~Shapiro,
  Phys.\ Rev.\ D {\bf 85} (2012) 104001
  [arXiv:1202.4235 [hep-th]].
\bibitem{RGE}
  E.~Elizalde and S.~D.~Odintsov,
  Phys.\ Lett.\ B {\bf 303} (1993) 240
   [Russ.\ Phys.\ J.\  {\bf 37} (1994) 25]  
  [hep-th/9302074]; Phys.\ Lett.\ B {\bf 321} (1994) 199
  [hep-th/9311087].


\bibitem{ParkerFulling}L.~Parker and S.~A.~Fulling,
  Phys.\ Rev.\ D {\bf 9} (1974) 341;
S.~A.~Fulling and L.~Parker,
  Annals Phys.\  {\bf 87} (1974) 176.

\bibitem{Ringwald}
  A.~Ringwald,
  Annals Phys.\  {\bf 177} (1987) 129
 \bibitem{Hu}S.~Sinha and B.~L.~Hu,
  Phys.\ Rev.\ D {\bf 38} (1988) 2423.
\bibitem{HuangBianchi}
  W.~H.~Huang,
  Class.\ Quant.\ Grav.\  {\bf 10} (1993) 2021
  [gr-qc/0401046].
\bibitem{Huanginho} 
 W.~H.~Huang,
  Class.\ Quant.\ Grav.\  {\bf 8} (1991) 83;
  Phys.\ Rev.\ D {\bf 48}, 3914 (1993)
\bibitem{Albareti}
  F.~D.~Albareti, J.~A.~R.~Cembranos and A.~L.~Maroto,
Phys. Rev. D {\bf 90}  (2014) 123509
  [arXiv:1404.5946 [gr-qc]] and Int. J. Mod. Phys. D (2014) 1442019
  [arXiv:1405.3900 [gr-qc]]
\bibitem{Schwarzschild} 
  P.~O.~Kazinski,
  Phys.\ Rev.\ D {\bf 80} (2009) 124020
  [arXiv:0909.3048 [gr-qc]].



\bibitem{Birrell} N.D. Birrell and P.C.W. Davies, {\it Quantum fields in curved space}, Cambridge (1982).

\bibitem{ColemanWeinberg}
  S.~R.~Coleman and E.~J.~Weinberg,
  Phys.\ Rev.\ D {\bf 7} (1973) 1888.
\bibitem{osborn}
  H.~Osborn,
  Nucl.\ Phys.\ B {\bf 363} (1991) 486;
I.~Jack and H.~Osborn,
  Nucl.\ Phys.\ B {\bf 883} (2014) 425
  [arXiv:1312.0428 [hep-th]].

\bibitem{slip}S.~F.~Daniel, R.~R.~Caldwell, A.~Cooray and A.~Melchiorri,
  Phys.\ Rev.\ D {\bf 77} (2008) 103513
  [arXiv:0802.1068 [astro-ph]].
\bibitem{Will}
C.~M.~Will,
  Living Rev.\ Rel.\  {\bf 17} (2014) 4
  [arXiv:1403.7377 [gr-qc]].
\bibitem{Amendola}
L.~Amendola, S.~Fogli, A.~Guarnizo, M.~Kunz and A.~Vollmer,
  Phys.\ Rev.\ D {\bf 89} (2014) 063538
  [arXiv:1311.4765 [astro-ph.CO]].
\bibitem{Ballesteros}G.~Ballesteros, L.~Hollenstein, R.~K.~Jain and M.~Kunz,
  JCAP {\bf 1205} (2012) 038
  [arXiv:1112.4837 [astro-ph.CO]].
\bibitem{Saltas}I.~D.~Saltas, I.~Sawicki, L.~Amendola and M.~Kunz,
  arXiv:1406.7139 [astro-ph.CO].
 
\bibitem{Uzan}
  J.~P.~Uzan,
  Living Rev.\ Rel.\  {\bf 14} (2011) 2
  [arXiv:1009.5514 [astro-ph.CO]].
\bibitem{Bertotti}
  B.~Bertotti, L.~Iess and P.~Tortora,
  Nature {\bf 425} (2003) 374.

\end{thebibliography}
\end{document}